\documentclass{aastex62}
\usepackage{subfigure}
\usepackage{graphicx}

\newcommand{\speed}[1]{#1 km~s${}^{-1}$}

\newcommand{\nfig}[1]{Figure~\ref{#1}}

\shorttitle{SOLAR FLUX ROPE ERUPTION}
\shortauthors{Chen et al.}

\begin{document}
\title{Observational Analysis on the Early Evolution of a CME Flux-rope: \\Pre-flare reconnection and Flux-rope's Footpoint Drift}

\correspondingauthor{Hechao Chen}
\email{chc@ynao.ac.cn}

\author[0000-0001-7866-4358]{Hechao Chen}
\affil{Yunnan Observatories,Chinese Academy of Sciences, 396 Yangfangwang, Guandu District, Kunming, 650216, China}
\affil{University of Chinese Academy of Sciences, 19A Yuquan Road, Shijingshan District, Beijing 100049, China}
\affil{Center for Astronomical Mega-Science, Chinese Academy of Sciences, 20A Datun Road, Chaoyang District, Beijing, 100012, China}

\author{Jiayan Yang}
\affil{Yunnan Observatories,Chinese Academy of Sciences, 396 Yangfangwang, Guandu District, Kunming, 650216, China}
\affil{Center for Astronomical Mega-Science, Chinese Academy of Sciences, 20A Datun Road, Chaoyang District, Beijing, 100012, China}

\author{Kaifan Ji}
\affil{Yunnan Observatories,Chinese Academy of Sciences, 396 Yangfangwang, Guandu District, Kunming, 650216, China}
\affil{Center for Astronomical Mega-Science, Chinese Academy of Sciences, 20A Datun Road, Chaoyang District, Beijing, 100012, China}

\author{Yadan Duan}
\affiliation{Yunnan Normal University, Department of Physics, Kunming 650500, Yunnan, China}


\begin{abstract}
We study the early evolution of a hot-channel-like magnetic flux rope (MFR) toward eruption. Combining with imaging observation and magnetic field extrapolation, we find that the hot channel possibly originated from a pre-existing seed MFR with a hyperbolic flux tube (HFT). In the precursor phase, three-dimensional tether-cutting reconnection at the HFT is most likely resulting in the heating and buildup of the hot channel. In this process, the forming hot channel was rapidly enlarged at its spatial size and slipped its feet to two remote positions. Afterward, it instantly erupted outwards with an exponential acceleration, leaving two core dimmings near its feet. We suggest that pre-flare reconnection at the HFT played a crucial role in enlarging the seed MFR and facilitating the onset of its final solar eruption. Moreover, a recently predicted drifting of MFR's footpoints was detected at both core dimmings. In particular, we find that MFR's west footpoint drift was induced by a new reconnection geometry among the erupting MFR's leg and thereby inclined arcades. As MFR's west footpoints gradually drifted to a new position, a set of newborn atypical flare loops connected into the west core dimming, causing a rapid decrease of dimmed area inside this core dimming and also generating a secondary flare ribbon at their remote feet. This reveals that core dimmings may suffer a pronounced diminishment due to the eruptive MFR's footpoint drift, implying that mapping the real footpoints of the erupting MFR down to the Sun's surface is more difficult than previously thought.
\end{abstract}

\keywords{Solar activity (1475), Solar coronal mass ejections (310), Solar flares (1496), Solar magnetic reconnection (1504)}

\section{Introduction} \label{sec:intro}
Solar eruptions are well-known drivers for most of extreme energetic space weathers in the heliosphere. They often simultaneously manifest as solar flares, filament eruptions, and coronal mass ejections (CMEs) at different spatial heights \citep{2011LRSP....8....1C,2015SoPh..290.3457S}. Due to their potential hazardous impacts on the near-Earth environment \citep[e.g.,][]{1991JGR....96.7831G,2018ApJ...861...28S} and significant disturbance at the global solar atmosphere \citep[e.g.,][]{1998GeoRL..25.2465T,2012ApJ...752L..23S}, solar eruptions have been intensively studied over the past decades, and succinctly depicted by a standard eruption scenario, i.e., the ``CSHKP" model \citep{1964NASSP..50..451C,1966Natur.211..695S,1974SoPh...34..323H,1976SoPh...50...85K}. In this model, the core magnetic field of solar eruptions, namely a magnetic flux rope (MFR), is simplified as a detached two-dimensional (2D) plasmoid \citep{1991ApJ...373..294F}; in the wake of the rising MFR, ``flare reconnection" is believed to be induced between two legs of overlying arcades within a vertical current sheet (see \citet{2003NewAR..47...53L} and therein references). As a result, the magnetically unleashed MFR would rapidly erupt upwards and blow out the overlying field to drive a CME \citep{1995ApJ...451L..83S}, while charged particles would impact downwards along reconnected field lines, resulting in a canonical ``two-ribbon'' solar flare in the chromosphere \citep{2003ApJ...598..683D,2011SSRv..159...19F}.
Despite this standard 2D scenario has made a great success in explaining the basic observational features of the CME/flare eruptive systems, it alone fails to reproduce other important aspects of the real three-dimensional (3D) solar eruption, including precursor activities that may facilitate the loss-of-equilibrium of eruptive MFRs, the complex distribution and dynamics of chromospheric flare ribbons (FRs), the physical information of eruptive MFR's footpoints, and so on.
\par
Generally, solar eruptions release their pre-stored magnetic energy via three phases, namely the precursor, impulsive, and gradual phases \citep{2001ApJ...559..452Z,2016ApJ...823L..19Z}. Thereinto, the latter two, jointly termed as the main phase, correspond to the impulsive acceleration of the erupting CME flux-rope, while the less-studied precursor phase includes key information on the eruptive structure and its trigger process. In the past, the limited observations showed that precursor activities of solar eruptions can come in various forms, in which preceding flux emergence \citep[e.g.,][]{2015A&A...583A..47P,2017ApJ...845...18Y,2019ApJ...874...96Y} or cancellation \citep[e.g.,][]{2011A&A...526A...2G,2016ApJ...816...41Y,2018ApJ...869...78C} in magnetograms, pre-eruption brightenings in H$_{\alpha}$/EUV images \citep[e.g.,][]{2012ApJ...758...42B,2016ApJ...823...41D,2017NatAs...1E..85W,2018ApJ...857..124A,2018ApJ...869...78C}, as well as non-thermal processes in microwave or hard X-ray wavelength \citep{2011ApJ...743..195J,2012ApJ...758..138A,2016ApJ...820L..37C} are most common ones. In recent years, in the precursor phase of many CME/flare eruptive events, a type of new progenitors of CME flux-ropes, namely hot channels \citep[HCs,][]{2011ApJ...732L..25C,2012NatCo...3..747Z,2013ApJ...763...43C}, have been detected in the AIA high-temperature passbands (e.g., 131 and 94 \AA). These newfound features commonly exist in the low corona with obvious helical/twisted fine fields or distinct writhed elbows \citep{2013ApJ...763...43C,2015A&A...580A...2Z}, and they are found to directly related to the occurrence of CMEs \citep{2013A&A...552L..11L,2013ApJ...764..125P} and major eruptive flares \citep{2015ApJ...808..117N}. To date, HCs have been evidenced as MFRs by some joint remote-sensing and in-situ observations \citep[e.g.,][]{2015ApJ...808L..15S}, but for many HC eruption events, two problems still remain elusive: (1) whether a corresponding MFR already exists before the HC eruption or is newly/partially built up during the eruption? (2) whether its loss-of-equilibrium is initially facilitated by the pre-flare reconnection beneath/above the HC \citep{1999ApJ...510..485A,2001ApJ...552..833M} or directly triggered the ideal magnetohydrodynamic (MHD) instabilities \citep{2004A&A...413L..27T,2006PhRvL..96y5002K}?
\par
Two-ribbon chromospheric flares had long been one basic module for building solar eruption models, but their unique 3D nature has received increasing attentions only in the recent decade. Based on previous observations, the 3D nature of two-ribbon flare can be described as two basic aspects. Regarding to spatial distribution, two-ribbon flares often demonstrate a unique but also common double J-shaped morphology on the either side of the main polarity inverse line (PIL) \citep[e.g.,][]{2014ApJ...788...60J,2016ApJS..225...16C}. Regarding to dynamics evolution, two-ribbon flares usually form and evolve with a distinct elongation-to-expansion dynamics \citep[e.g.,][]{1990SoPh..125..321K,2009ApJ...692.1110Q,2017SoPh..292...25P}, although some of them may display more complex elongation patterns sometimes \citep{2009ApJ...690..347L}. Recently, \citet{2012A&A...543A.110A,2013A&A...549A..66A} and \citet{2013A&A...555A..77J} have extended the standard eruption model into three dimensions via 3D dynamical MHD simulations. In this model, the MFR is anchored at both ends and wrapped by a thin 3D quasi-separatrix layer \citep[QSL,][]{1996A&A...308..643D,1997A&A...325..305D} where the magnetic field connectivity remains continuous but changes with a sharp gradient \citep{2002JGRA..107.1164T,2012A&A...541A..78P}. Accordingly, flare reconnection is predicted to occur at a hyperbolic flux tube \citep[HFT,][]{2002JGRA..107.1164T} where the MFR-related QSL crosses itself under the MFR with the highest connectivity change, while two-ribbon flare would well distribute along the double J-shaped photospheric footprint of the MFR-related QSL \citep{2014ApJ...788...60J,2015ApJ...810...96S,2016ApJ...817...43S}. In particular, if the erupting CME flux-rope acquires a higher twist, the umbrella-like extremes of FRs may even close on themselves, forming closed hooks and enclosing the CME flux-rope's feet \citep{1996JGR...101.7631D,2012A&A...541A..78P,2016ApJ...823...62Z}.
\par
In addition to the typical two-ribbon flares, a statistical study from \citet{2014ApJ...782L..27Z} shows that 11 of 19 X-class flares happened with secondary FRs. Compared with the main two-ribbon flares, these atypical secondary FRs are not connected by post-flare loops and appear in a weaker emission. Most of these secondary FRs appeared simultaneously with the two-ribbon flares, but others may appear later than the two-ribbon flares. These newfound complex FRs seem to be triggered via 3D reconnection process, but how they couple with the main eruption has not been fully understood. To date, most observational works of multiple-ribbon flares commonly contributes their generation to the existence of a complex flux distribution \citep{2015ApJ...809...45Z,2016ApJ...823..136Z} or a null-point type configuration \citep{2014ApJ...781L..23W,2017ApJ...838..134B} over the eruption region. In particular, \citet{2015ApJ...812...50J,2017ApJ...845...26J} have conducted a series of observational analysis on two intriguing ``three-ribbon'' events that occurred in large-scale fan-spine-type magnetic configurations. They found that the reconnection occurring in the wake of the erupting sigmoid produces the parallel two-ribbon flare, while the ensuing null-point reconnection resulted in the quasi-circular secondary FR just like the situation that occurs in small-scale fan-spine jet events \citep[e.g.,][]{2016NatCo...711522J,2019ApJ...872...87L}.
\par
Coronal dimmings are transient dark regions that are distinctly observed in the early eruption phase of CMEs with EUV, X-ray, and even H$_{\alpha}$ passbands \citep{1997ApJ...491L..55S,1998GeoRL..25.2465T,1999ApJ...520L.139Z,2003ApJ...597L.161J}. They have been divided into core and secondary dimmings \citep{2019ApJ...874..123D,2019ApJ...879...85V}. Different from the irregular, widespread, and short-lived secondary dimmings, the long-lived core dimmings have a deeper and faster density reduction \citep{2018ApJ...857...62V}, and often conjugately present at the extremes of FRs, sometimes enclosed by flaring hooks \citep{2016ApJ...817...43S,2019arXiv190802082Z}. Thus, core dimmings has been explained as the line-tied footpoints of the evacuated CME flux-rope \citep{2000JGR...10527251W,2007SoPh..240...77J,2018ApJ...857...62V,2019ApJ...879...85V}.
Focusing on this conjecture, \citet{2019A&A...621A..72A} recently also analysis the physical evolution of the CME flux-rope's footpoints under the framework of the standard solar eruption model in 3D \citep{2005A&A...430.1067A}. Via identifying and tracing 3D reconnecting field lines with time, however, they found that CME flux-rope's footpoints can undergo an unexpected drifting due to a new reconnection geometry among flux rope's leg and thereby inclined arcades, namely ``ar-fr" reconnection. Meanwhile, the erupting CME flux-rope would be either eroded on its inner side and also enlarged on the external side. As a result, their 3D simulation inversely argues that core dimmings can not be curtly used to locate the footpoints of the erupting CME flux-rope or even the pre-eruptive MFR. More recently, the predicted footpoint drift of the erupting MFR has been evidenced by direct observations \citep{2019arXiv190802082Z,2019arXiv190903825L}. But whether the footpoint drift of erupting CME flux-rope can pronouncedly influence the evolution of core dimmings, including its dimmed area and lifetime, are calling for new observations.
\par
In attempt to shed light on these issues, in this paper, we present an observational analysis on the early evolution of a hot-channel-like CME flux-rope, focusing on its clear pre-flare activities, flux-rope's footpoint drift, as well as its associated secondary FR. Accordingly, we discussed the possible origin/initiation process of the hot channel, and investigated the generation of the secondary FR. Moreover, we also find that the drifting of MFR¡¯s footpoints can directly lead to a rapid diminishment of core dimmings, supporting that core dimmings might can not accurately map the erupting CME flux-rope footpoints in real 3D solar eruptions. The layout of the paper is as follows: In Section 2, we describe the instruments and data. Our main observational findings are shown in Section 3. In Section 4, we summarize these results, and further discuss their implications.
\section{Instruments and Data}
The observational data we used are provided by the Atmospheric Imaging Assembly \citep[AIA,][]{2012SoPh..275...17L} and Helioseismic and Magnetic Imager instruments \citep[HMI,][]{2014SoPh..289.3483H} on board
\textit{Solar Dynamic Observatory} \citep[\textit{SDO,}][]{2012SoPh..275....3P}, as well as the \textit{New Vacuum Solar Telescope} \citep[NVST,][]{2014RAA....14..705L} that operated in \textit{Fuxian Solar Observatory} in China. The AIA provides the full-disk EUV/UV imaging observations of the solar atmosphere from the photosphere up to the corona through 10 passbands, with a pixel size of 0.$^{\prime\prime}$6, high temporal cadence (12 s), and wide temperature range (0.06$-$20 MK); the HMI measures the full-disk photospheric magnetic fields at 6173 \AA, providing routine line-of-sight (LOS) and vector magnetograms with a pixel size of 0.$^{\prime\prime}$5. We choose 131, 94, 211 and 193 \AA \ images to analysis the detailed formation and eruption of the hot channel, and apply 304 and 1600 \AA \ images to study its related chromospheric FRs.
Meanwhile, the HMI LOS magnetograms are used to analysis the preceding evolution of photospheric magnetic fields. The NVST is designed to observe the Sun with high temporal (12 s) and spatial (0.$^{\prime\prime}$165) resolutions, and now it images the solar chromosphere at the emission line of H$_\alpha$ 6562.8 \AA \ with a bandwidth of 0.25 \AA \ . Its raw data are first calibrated by subtracting dark currents and flat field, and then reconstructed to Level 1$+$ using a speckle masking method \citep{2016NewA...49....8X}. On 2014 March 20, NVST focused on the NOAA active region (AR) 12010 with a field of view (FOV) of $180^{\prime\prime}\times 180^{\prime\prime}$, thus respectively covers the pre-flare (01:24:45-01:53:13 UT) and decay phase (03:54:10-05:40:55 UT) of this flare event. Accordingly, we adopted the Level 1$+$ NVST H$_\alpha$ observations to investigate the related chromospheric responses before and during the flare, and also to present the rapid diminishment of the west core dimming. All NVST images are mapped to the helioprojective-cartesian coordinates aided by a new automatic alignment technique \citep[See details in][]{2019CSB...64...16J}. To compensate for solar rotation, the SDO images taken at different times are aligned to an appropriate reference time.
\par
To check the pre-flare magnetic configuration of the solar eruptions, we also choose the vector magnetogram for extrapolation at 03:12 UT from the HMI Active Region Patches (SHARPs) product \citep{2014SoPh..289.3549B}. The vector magnetograms are computed by the Very Fast Inversion of the Stokes Vector code \citep{2011SoPh..273..267B}, and the remaining 180$^{\circ}$ ambiguity in transverse component has been resolved with the Minimum Energy algorithms \citep{1994SoPh..155..235M,2009SoPh..260...83L}.  Before being taken as the bottom boundary, the vector magnetogram is ``pre-processed"  to best suit the force-free condition \citep{2006SoPh..233..215W}. Using the NLFFF package available in Solar SoftWare (SSW)\footnote{\url{http://www.lmsal.com/solarsoft/ssw_packages_info.html}}, we build an NLFFF model that approximates the 3D coronal field with the ``weighted optimization'' method \citep{2000ApJ...540.1150W,2004SoPh..219...87W}.
Our calculation is performed within a box of 176 $\times$ 144 $\times$ 144 uniformly spaced grid points, which corresponds to about 64 $\times$ 52 $\times$ 52 Mm$^3$ and well encloses the eruption source region. Moreover, to better identify the special topological features (including MFRs, QSLs and HFTs), we also compute the squashing factor (Q) and twist number ($T_{w}$) of the extrapolated field utilizing a topological analysis code that developed by \citet{2016ApJ...818..148L}.
\section{Observational Results}
\subsection{Overview}
\begin{figure}[htb!]    
   \centerline{\includegraphics[width=0.8\textwidth,clip=]{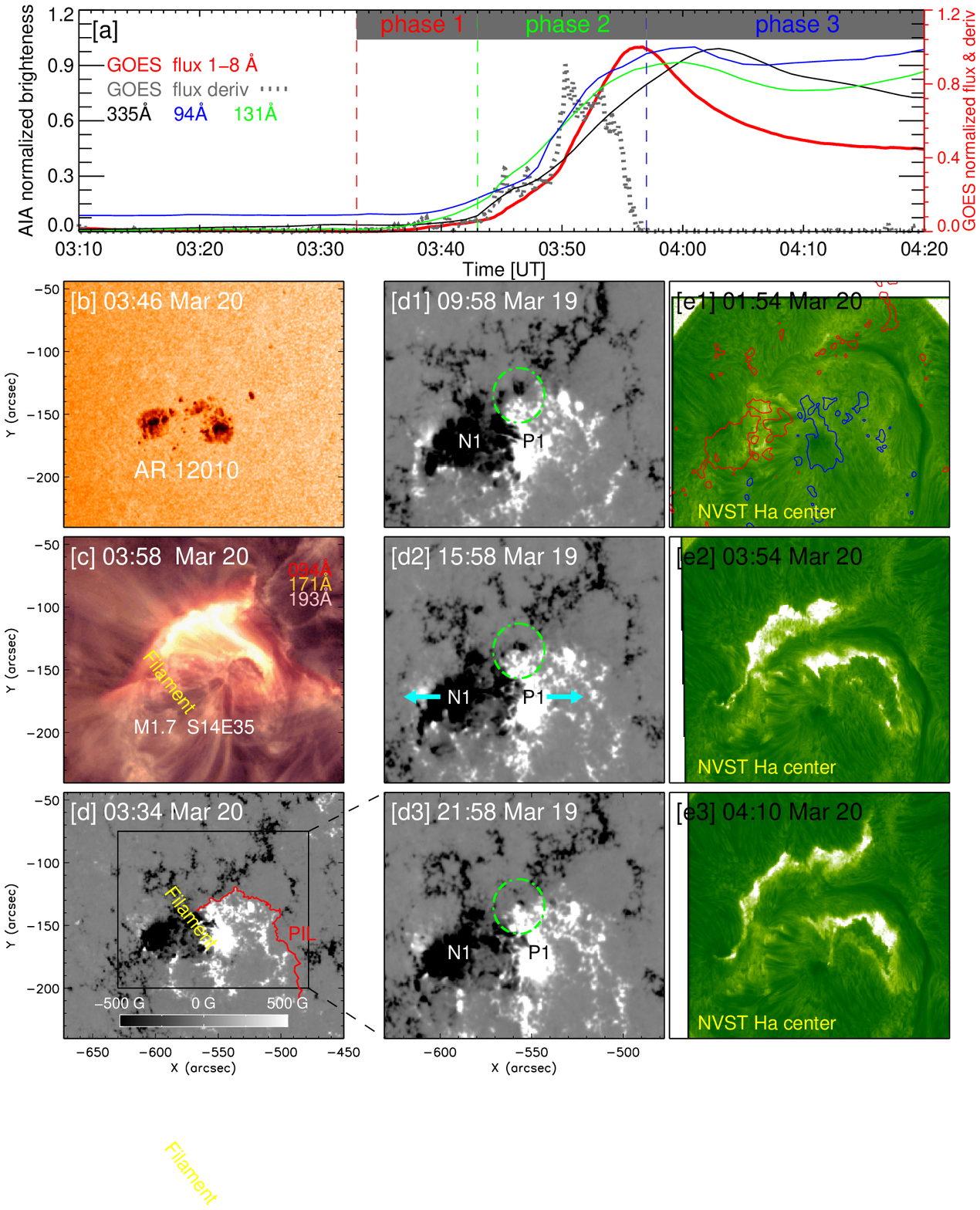}
              }
              \caption{(a) The normalized AIA light curves that calculated in the field-of-view (FOV) of panel (c), as well as the variation of normalized \textit{GOES} soft X-ray flux at 1-8 \AA \ and its derivative. (b) HMI intensity image, (c) the composite AIA image, and (d) line of sight magnetogram show the general configuration of the AR 12010, with red line denoting the main PIL. (d1-d3) The pre-eruption photospheric magnetic evolution, with the green dash circle enclosing the flux cancellation region. The cyan arrows in (d2) indicate the expanding motion of main sunspot, while $N1$ and $P1$ denote their negative/postive flux. (e1-e3) NVST $H_\alpha$ images present the chromospheric response before and during the flare, in which a curve filament stays intact from the eruptive event. In panel (e1), the red/blue contour denotes negative/positive flux with a saturate value at $\pm$ 500 G.}
   \label{fig1}
\end{figure}

The eruption event under study occurred at the AR 12010 on 2014 March 20 (see \nfig{fig1}(b-c)), which simultaneously associated with a M1.7 solar flare and a CME.
According to AIA fluxes and the GOES SXR 1-8 \AA \ flux (see \nfig{fig1}(a)), the M1.7 flare started, peaked, and ended at 03:33, 03:56, and  04:08 UT, respectively.
\nfig{fig1}(d)-(d3) show the overall photospheric evolution of the AR 12010. It consists of two opposite-polarity sunspot ($N1$ and $P1$) with a $\beta\gamma$ field configuration (see \nfig{fig1}(b)). At the north border of the active region, a polarity inversion line (PIL) exists among the positive flux ($P1$) and other background negative fluxes. After 09:58 UT on March 19, episodes of slow flux cancellation started near the east section of the PIL and lasted for more than 12 hrs (marked by green dash circles in \nfig{fig1}(d1)-(d3)).
Following the preceding flux cancellation, subsequently, the buildup and inflation of a hot-channel-like solar MFR toward eruption, and multiple chromospheric FRs, as well as the rapid diminishment of core dimmings were well captured by the AIA above the east section of the PIL.
Note that along the PIL, the NVST H$\alpha$ images show that a long filament remained intact through out the whole eruption process (see the left column of \nfig{fig2}). Considering that this filament did not show any obvious interaction to the hot channel eruption, we tend to deal it as an unconcerned observed feature through out this study.
\subsection{The buildup and eruption of the hot-channel-like MFR via two-stage reconnection}
\nfig{fig2} presents the buildup of the hot channel in 131 \AA \ images during the precursor phase (03:30$-$03:43 UT) (see also its animation). Following the above-mentioned photospheric flux cancellation, a preceding plasma heating signal was first captured right above the canceling side in 1600 and 131 \AA \ passbands. Afterwards, some arcade-like structures started to be heated from the background near the same location. By about 03:35 UT, a set of short flaring loops formed below these rising bright arcades, indicating magnetic reconnection took place there. With the reconnection going on, the arcade-like structure subsequently extended upward with an enhanced emission, and gradually took on a twisted shaped with a clear elbow near its east end. This distinct twisted morphology (as outlined by the black dash line in panel (b5)) suggests that the hot channel should possess some degree of magnetic twist, corresponding to a forming MFR \citep{2013ApJ...763...43C}. Note that the newborn hot channel only can be unambiguously observed in 131 \AA \ images, suggesting plasma in the hot channel has also been heated to near 10 MK during its formation process. Below the forming hot channel, short flaring loops underwent a bidirectional movement after their appearance, forming a neat row of sheared flare loops.
As the footpoints of reconnected field lines, in the simultaneous 304 \AA \ images (the left column of \nfig{fig2}), sporadic flare kernels first appeared at 03:35 UT, and then elongated along the PIL forming a skeleton of the two-ribbon main flare in the chromosphere.
\par
As shown in the right part of \nfig{fig3}, soon after its rapid formation, the expanding hot channel instantly erupted towards the southeast by the time of about 03:44 UT. Accordingly, in the running-difference images in 131 and 211 \AA \ images, the eruptive hot channel soon developed as a nascent CME at the low corona. It is worthy noting that the nascent CME consists of an inner core and a weak bubble-like shell. The composite image for 94, 131, and 211 \AA \ in \nfig{fig4}(a) also reveals that its inner core corresponds to a high-temperature eruptive MFR, and its shell corresponds to a low-temperature leading edge ($LE$). Possibly due to the absent of cool plasma \citep{2011ApJ...732L..25C}, the bright inner core of the CME that observed in 131 \AA \ images thus inversely demonstrates as a dark cavity in simultaneous 211 \AA \ images (see \nfig{fig3}(c1) and (d1)). By the time of 05:12 UT, an corresponding CME was detected in the coronagraph observations from Large Angle and Spectrometric Coronagraph aboard the {\sl Solar and Heliospheric Observatory}. As its chromospheric response, the newly-formed two flare ribbons underwent an obvious lateral expansion movement perpendicular to the PIL (see the left column of \nfig{fig3}). Meanwhile, two core dimmings conjugately appeared at the extremes of the two FRs \citep{2007ApJ...662L.131J}, roughly encircled by flaring hooks.
\par
By tracking the $LE$ and the eruptive MFR along the slice $S1$ in \nfig{fig4} (a), we obtain their height-time plots and velocity profiles (\nfig{fig4}(b) and (c)). From which, one find that the hot-channel-like MFR first underwent a slow rise before its final eruption, while the $LE$ later formed ahead of the erupting MFR. The projected velocity of the MFR and its $LE$ increased up to around 110 and  \speed{350} during 03:40-03:50 UT, respectively. Apparently, their accelerations are well correlated with the impulsive phase of the eruptive flare \citep{2001ApJ...559..452Z}. Using an edge detector algorithm, we also quantify the expansion of two-ribbon flare along the slice $S2$ in \nfig{fig5}(d). As shown in \nfig{fig4}(e) and (f), the two-ribbon flare initially appeared in the precursor phase (at $\sim$ 03:42 UT), and its skeleton subsequently expanded its distance from 7 to 35 Mm at with a velocity up to \speed{20} in the impulsive phase. Such a distinct elongation-to-expansion transition in FR dynamics can also be clearly distinguished by the running-difference 304 \AA \ image in \nfig{fig4}(d), supporting that two-stage magnetic reconnection, i.e., 3D pre-flare reconnection and quasi-2D ``flare reconnection", respectively participated the buildup and eruption of the hot channel with time.
\subsection{NLFFF extrapolation and topology analysis of pre-eruptive magnetic configuration}
Although imaging observations support that the hot channel rapidly formed before its eruption, whether it originates from a seed  MFR or newly formed via reconnection is unconfirmed. To answer this question, the perspective views of the extrapolated  pre-eruptive coronal magnetic configuration and related QSL maps are shown in \nfig{fig5}. Through mapping magnetic connectivities and computing the twist number $T_{w}$, a coherent MFR is well identified (as green lines) right above the east section of the PIL where the main FRs appeared. This pre-existing MFR resides in around 2.35$-$6.70 Mm above the photosphere with a $T_{w}$ range from 0.96$-$1.43. Correspondingly, its average and median $T_{w}$ are $\sim$ 1.24 and 1.38. Meanwhile, two set of less-sheared enveloping fields, $L1$ and $L2$ , are found to surround this twisted MFR, which are traced by blue and red lines. $L1$ and $L2$ take on a arcade shape and root their footpoints at opposite-polarity fluxes below the middle section of the MFR. Such an unique configurations have been revealed in several previous observations \citep{2010ApJ...725L..84L,2013ApJ...778L..36L,2018ApJ...869...78C}, which is thought to be in favor of the occurrence of tether-cutting reconnection \citep{2001ApJ...552..833M,2010ApJ...708..314A}.
\par
Interestingly, in \nfig{fig5}(d), a vertical Q-factor cross section that perpendicular to the pre-existing MFR reveals that a typical hyperbolic tube (HFT) existed in the core of the MFR-related QSL prior to the eruption \citep{2012ApJ...750...15S,2018ApJ...868...59L}. Accordingly, consistent with the result of \citet{2016ApJ...823...62Z}, the pre-existing MFR in our case also resides above the HFT surrounded by a teardrop QSL volume (white dashed line), while sheared arcades ($L1$ and $L2$) reside at right or left of the HFT. Based on previous simulation and observations, the pre-flare reconnection in our event should happen at the HFT, because it is the privileged place both for current layer build-up and the occurrence of 3D reconnection \citep{2012ApJ...744...78S,2014ApJ...788...60J}. In that case, the newborn hot channel is most likely originate from the seed MFR, and further enlarged its spatial size and obtained a hot-temperature property via a tether-cutting type 3D pre-flare reconnection.
\par
To compare the FR distribution and its related QSLs, we overlaid the high-Q tracers (with $log$(Q)$> 4$) at Z = 1) above the photosphere on a negative 304 \AA \ flare image in \nfig{fig5}(c). It can be seen that most parts of the double J-shaped FRs, including two straight sections ($S+$ and $S-$) and its east hook ($H-$), are well matched by hight-Q tracers; however, the west hook ($H+$) seemingly deviates the high-Q tracers. Likewise, the foot locations of the pre-existing MFR are also overlaid in \nfig{fig5}(c) and marked by cyan diamonds. Apparently, the pre-existing MFR's feet located at the straight FRs, instead of the hooked FR extremes where the final erupting hot channel might rooted its feet. Consistent with the predictions that proposed by the recent 3D extensions to the standard eruption model, these results support that main FRs did form along footprints of the main QSLs \citep{2012A&A...543A.110A,2013A&A...555A..77J,2015ApJ...810...96S}.
In addition, to understand the stability of the pre-existing MFR, we also obtain the overlying horizontal field distribution along the while dash line in \nfig{fig5}(c) via computing magnetic decay index, i.e, $n = -\frac{\partial ln(B_h)}{\partial ln(h)}$ \citep{2006PhRvL..96y5002K}. The result as seen in \nfig{fig5}(e) indicates that the critical index (1.5) only appears in the domain where is higher than 12 Mm, implying that the pre-existing MFR can survive from the ideal torus instability unless it experiences an extra disturbance that lifts it into the higher torus-unstable domain. Note that the seed MFR is found to have an averaged $T_{w}$  $\sim$ 1.24 turns but still kelp itself kink-stable, we conjecture that it may possess a higher critical value of kink instability due to its slim configuration \citep{2006PhRvL..96y5002K,2016ApJ...818..148L}.
\subsection{The footpoint drift of the hot-channel-like MFR and its associated phenomena}
It is worthy noting that in this event, the forming hot-channel-like MFR continuously inflated or expanded itself before the final eruption (refer to \nfig{fig2}). In fact, the detailed observations indicate that such inflation or expansion actually corresponds to a dynamic enlarging process of the eruptive MFR in the precursor phase. Of particular interest is that the enlarging process of the eruptive MFR associates with obvious footpoint drifts at both of its legs. Accordingly, this footpoint drift process demonstrated as a typical slipping reconnection signals near the east feet of the MFR, and it also resulted in the generation of a secondary FR and a rapid diminishment of core dimming near the west feet of the MFR.
\subsubsection{Observational signals of MFR's footpoint drift and slipping reconnection}
To emphasize its related footpoint drift, here we briefly show the enlarging process in \nfig{fig6}. In terms of the morphology, the forming hot channel initially demonstrated as an flaring arch-like structure with a small spatial size at about 03:34 UT. But, by the time of 03:42 UT, the formed hot channel rapidly acquired a larger size and took on a complex morphology with a writhed elbow near its east feet. For its feet, a close inspection can find that both its feet also drifted to new positions. In panel (b), the two purple arrows respectively illustrate the rough drifting orientations of its east and west feet. Overall, consistent with the enlarging trend of the eruptive MFR, two MFR's feet both drifted towards outside. Moreover, to reveal the relationship between the two-ribbon flare and MFR's feet, the outlines of the newborn hot channel that obtained at three different moment have also been plotted in panel (c). It can be seen that from 03:34 UT to 03:48 UT, following the enlarging process of the hot channel, its feet both synchronously drifted along two flaring hooks near the extremes of the two-ribbon flare. Moreover, referring to the feet position of the extrapolated seed MFR in \nfig{fig5}(a-c) marked by cyan diamonds, one can also find that the core dimmings apparently can not map the footpoints of the pre-flare magnetic structure, too, due to the footpoint drift of the eruptive MFR.
\par
Focusing on the east feet of the eruptive MFR, the close relationship among the MFR's footpoint drift, the formation of flaring hook, and slipping flare kernels are also exhibited in \nfig{fig7} (and its animations). As shown in panels (a1-a2), following the buildup of the twisted field lines in the hot channel, the east feet of the hot channel gradually drifted into the center of the negative-polarity sunspot of AR 12010. Via a closed-up inspections of field lines that traced by heated plasma near its east elbow in the 131 \AA \ images, we noticed that the hot channel anchored its footpoints at a chain of tiny bright knots above the FR (as marked by white arrows in panel (a2). Interestingly, these bright knots displayed a slipping motion from the west towards the east during 03:37-03:42 UT. At the same time period, as shown in 304 \AA \ images, the FR's extreme also slipped with a average velocity range from 14 to \speed{67}, resulting in the closed FR hook. As the well-accepted observational signatures of 3D slipping reconnection \citep{2014ApJ...784..144D,2015ApJ...804L...8L}, the observed slipping behaviors of reconnected field lines and flare kernels further reinforce the idea that a 3D pre-flare reconnection was responsible for the formation and also the enlarging of the hot channel.
\subsubsection{The associated phenomena: secondary flare ribbon}
Near the west feet of the eruptive MFR, a semi-circular secondary FR was found to form outside the eruption site after the main flare reached its peak. As shown in \nfig{fig8} and its animation, this new FR first appeared and then fully formed at around 03:55 and 04:25 UT, respectively. From the northeast to the southwest, it elongated along an arc trajectory and reached to its maximum length $\sim$ 110 Mm.
Note that the emission of this secondary FR is no strong like two main ribbons. In particular, a set of atypical flare loops almost simultaneously appeared near the west leg of the eruptive hot channel, connecting to the newborn secondary FR (near the negative magnetic flux $N2$) and the positive-polarity hooked FR ($P2$) (see \nfig{fig8} (b)-(d)). Compared with the typical post-flare loops, these newborn flare loops were weaker in emission, and thus can only be better recognized in the running-difference 94 \AA \ images. We believe that the appearance of atypical flare loops and its related secondary FR suggest a new external reconnection during the 3D eruption process of the hot channel.
\par
The potential field extrapolation in \nfig{fig8}(e) shows that there is a set of pre-existing loops connecting $P1$ and $N$ before solar eruption commenced, which can also be evidenced by the observed loops in 171 \AA \ image (see \nfig{fig8}(f)). When the eruptive MFR erupted, the pre-existing loops were found to shrink towards the opposite-polarity MFR leg. Accordingly, a new reconnection may be gradually initiated among the shrinking loops and the west leg of the MFR. This could explain why the formation of secondary FR started from a proximal end to a distal end. Note that such a reconnection geometry is analogous with the configuration that required in the so-called ``ar-fr" reconnection \citep{2019A&A...621A..72A}. Via an inspection of 1600 \AA \ image subsequences, a related footpoint brightening signature was also found to support this scenario. As it can be seen in \nfig{fig8}(h)-(i), at the original footpoint position of the pre-existing loops ($P1$, marked by yellow circles), an obvious UV brightening indeed suddenly appeared following the appearance of the atypical flare loops by the time of 03:55 UT. As a result, part field lines of the eruptive MFR should drift their feet to $P1$ (yellow circles), and the pre-existing loops changed their positive-polarity footpoints to $P2$ leading to the the appearance of atypical flare loops and the secondary FR.
These observations thus tend to support that a new extra 3D reconnection simultaneously resulted in the west footpoint drift of the eruptive MFR and the generation of the semi-circular secondary FR in the late phase of the solar eruption.

\subsubsection{The associated phenomena: core dimming diminishment}
Due to the ongoing extra 3D reconnection, a rapid core dimming diminishment was also been detected near its original west feet during 03:55 UT to 04:30 UT. As shown in the top and middle rows of \nfig{fig9}, several selected NVST H$\alpha$ and running-difference 94 \AA \ images reveal the close relationship between the appearance of atypical flare loops and the diminishment of west core dimming. Consistent with the generation timing of the secondary FR, atypical flare loops suddenly formed and further heaped up near the original west leg of the erupting CME flux-rope after 03:55 UT.
As one of observational signals of MFR's footpoint drift, these atypical flare loops now gradually drifted their footpoints into the core dimming region. Accordingly, in NVST H$\alpha$ \AA \ images, the dimmed area in the west core dimming also underwent a rapid decrease at the same time period. Note that during this diminishment process, the dimmed pixels inside the core dimming were invaded by new brighten pixels, which is quite different from the natural disappearance of coronal dimmings that caused by a gradual plasma replenishment.
\par
To give a better description on this diminishment of the west core dimming, we also quantified the time variation of the integral magnetic flux that through the core dimming region, $\Phi_{DM}$. Considering that the west dimming region in 304 \AA \ images was least interfered by coronal structures and suffered less projection effect, we thus take it as the best identified subject. The dimmed area of the core dimming is detected utilizing several morphological algorithms. Through a trial-and-error approach, an optimal threshold can be set to best detect the pixels of the dimmed region (see \nfig{fig9}). Afterwards, we project all these AIA data to a related pre-flare CEA $B_r$ map with a same pixel size and computed $\Phi_{DM}$ over all identified pixels at different times. In addition, error bars for $\Phi_{DM}$ were given by performing similar identification and calculation with a varied detection thresholds ($\pm$ 10$\%$), respectively. As shown in \nfig{fig9}(d), the time variation of $\Phi_{DM}$ was obtained during 03:45 UT to 04:30 UT. From which, it is clear that $\Phi_{DM}$ first reached it peak ($1.70 \times 10^{20}$ Mx) at 03:58 UT, and then started to decrease as the atypical flare loops obviously formed at $\sim$ 04:05 UT, and finally drop to $0.65 \times 10^{20}$ Mx by the time of 04:30 UT. This result provides clear evidence to support that the footpoint drift of the erupting CME flux-rope indeed can pronouncedly alter the total area and lifetime of core dimming.
\section{Summary and Discussion}
Using multi-wavelength imaging observations from \textit{SDO}/AIA, HMI, and NVST, we have presented a detailed observational analysis on the early evolution of a hot-channel-like solar MFR towards eruption in AR 12010.
This event associates with multiple chromospheric FRs (including a two-ribbon flare  and a secondary FR) and clear signatures of MFR's footpoint drift, which provides us a good opportunity to study the inherent 3D eruptive process of CME flux-rope eruption. Combining with NLFFF extrapolation and topology analysis, we not only investigated the possible origin/initiation process of the hot channel, but also studied the distribution and dynamics of its related chromospheric flare ribbons. Moreover, a direct link among the predicted drifting of MFR's footpoints, the generation of secondary FR, and a rapid diminishment of core dimmings was built up. Here, key aspects of our observations are listed as follow.
\begin{enumerate}
\item{The NLFFF extrapolation and imaging analysis results support that the hot channel possibly originated from a pre-existing seed MFR with a HFT configuration. Accordingly, the pre-flare 3D reconnection is most likely triggered at the HFT, further resulting in the heating and enlarging of the hot-channel-like MFR.}
\item{From the precursor to impulsive phase, the main flare two ribbons underwent an unique elongation-to-expansion dynamics. This implies that the buildup and eruption of the hot channel seemingly accomplished with two-stage reconnection: the tether-cutting type pre-flare reconnection and an ensuing quasi-2D ``arcade reconnection".}
\item{During the buildup of the hot channel, obvious MFR's footpoint drift signatures were detected near two conjugate core dimmings. In particular, the drift of its east footpoints manifested as an apparent slipping motion of flare kernels, resulting in the formation of a closed FR hook.}
\item{Near the west core dimming, the footpoint drift of the erupting MFR was found to be caused by an extra reconnection among the feet of the eruptive hot channel and thereby inclined arcades. More interestingly, this new reconnection geometry further resulted in the generation of a semi-circular secondary FR and a set of atypical flare loops.}
\item{In addition, we find that in our event, MFR's footpoint drift pronouncedly decreased the dimmed area of the west core dimming. During this process, dimmed regions inside the west core dimming was rapidly invaded by new flaring region following the appearance of atypical flare loops.}
\end{enumerate}
\par
\begin{figure}[htb!]    
   \centerline{\includegraphics[width=1.0\textwidth,clip=]{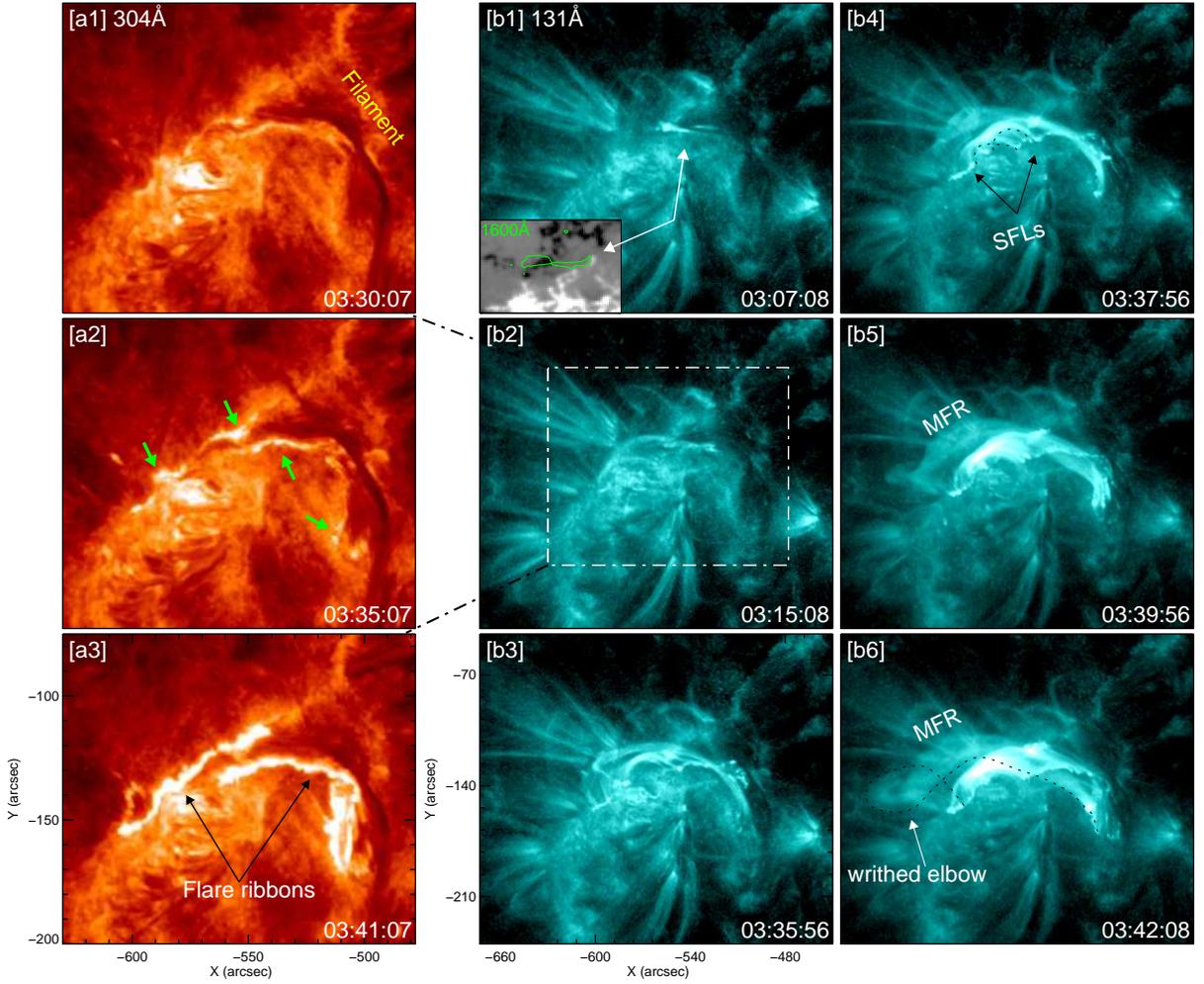}
              }
              \caption{(a1-a3): Selected 304 \AA \ images display the appearance of sporadic flare kernels and their elongation along the PIL. In panel (a2), green arrows denote the newborn flare kernels. (b1-b5): Sequential 131 \AA \ images present the rapid buildup of the hot-channel-like MFR. The insert in panel (b1) illustrates a contour of the preceding 1600 \AA \ brightening over the flux canceling site. In panel (b5), an obvious writhed elbow can be noticed at its east feet. The brown box in (b4) represents the FOV of \nfig{fig6}. An animation of panels (a2) and (b2) is available. The animation covers 03:25:07 UT to 03:43:07 UT with 60 s cadence. The video duration is 1 s.}
   \label{fig2}
\end{figure}
\begin{figure}[htb!]    
   \centerline{\includegraphics[width=1.0\textwidth,clip=]{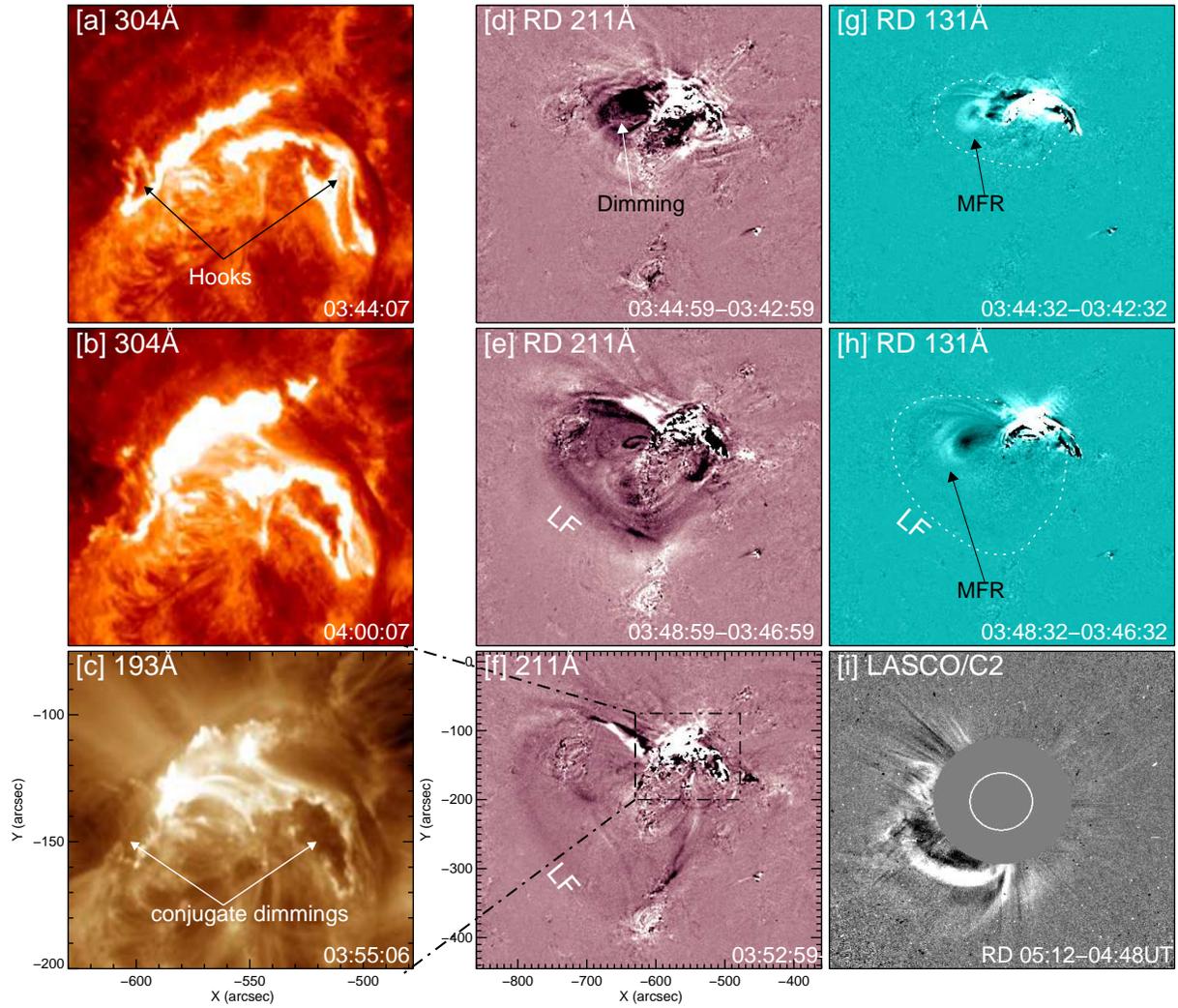}
              }
              \caption{(a)-(b): The lateral expansion of the main FRs observed in 304 \AA \ images. (c): The conjugate core dimmings that developed in 193 \AA \ image. (d-h): Running-difference 211 and 131 \AA \ images shows the formation and eruption of the nascent CME in the corona. In which, the hot core in 131 \AA \ images, namely the eruptive MFR, simultaneously displayed as a dimming core in 211 \AA \ images; ``LF" denotes the leading front that developed ahead the erupting MFR. (i): The corresponding CME that observed by LASCO/C2 in the higher corona. An animation of panels (a), (c), (d), and (g) is available. The animation covers 03:41:07 UT to 03:56:07 UT with 60 s cadence. The video duration is 1 s.}
   \label{fig3}
   \end{figure}

For many hot channel eruption events, their exact origin and initiation process still remain heated debated yet. Some observation supported that before the hot channel eruption, a corresponding MFR may already exist \citep[e.g.,][]{2016ApJ...818..148L,2016ApJ...817..156W,2019ApJ...878...38Y} and slow photospheric flows might be important for its gradual formation \citep[e.g.,][]{2018ApJ...856...79Y,2018A&A...619A.100H,2019ApJ...874..182V};
while others argued that they can also be newly built up via rapid coronal reconnection during the eruption \citep{2011ApJ...732L..25C,2014ApJ...792L..40S,2015A&A...583A..47P}. For their loss-of-equilibrium, some researchers found that the loss-of-equilibrium of hot channel is initially facilitated by the breakout type \citep[e.g.,][]{2012ApJ...750...12S,2016ApJ...820L..37C,2018ApJ...869...69M} or tether-cutting type  \citep[e.g.,][]{2014ApJ...797L..15C,2015ApJ...812...50J,2016ApJ...823...41D,2018ApJ...869...78C} pre-flare reconnection, but others believed the loss-of-equilibrium of hot channel may be directly triggered the ideal MHD instabilities \citep[e.g.,][]{2007ApJ...661.1260L,2015ApJ...805...48B,2017ApJ...850...38V}.
Note that \citet{2019arXiv190808643D} recently carried out a survey on the initiation mechanism of all the major solar flares, either eruptive or confined, from 2011 to 2017. Statistically, their study results indicates that magnetic reconnection and ideal MHD instabilities seemingly play an equal role in the triggering of major flares and eruptions. In the present study, our NLFFF extrapolation and topology analysis shows that a seed MFR existed within a HFT configuration prior to the later solar eruption. Near the HFT, two sets of sheared arcades serve as enveloping field well holding the seed MFR, and rooted their opposite footpoints at canceling fluxes. Combining these results with imaging observations, we suggest that the pre-flare 3D reconnection is most likely first triggered at the HFT in the precursor phase due to the preceding flux cancellation \citep{2012ApJ...750...15S}, and the newborn hot channel should originate from the pre-existing seed MFR. Through pre-flare 3D reconnection, the seed MFR rapidly be heated up to 10 Mk manifesting as a newborn hot channel, and its spatial size also underwent an enlarging process at the same time (as revealed by \nfig{fig2} and \nfig{fig5}). As a result, the enlarging hot channel may readily lift into the above-mentioned torus-unstable domain (around 12 Mm above the photosphere), and soon enter its exponential accelerated eruption.
\par
\begin{figure}[htb!]    
   \centerline{\includegraphics[width=1.2\textwidth,clip=]{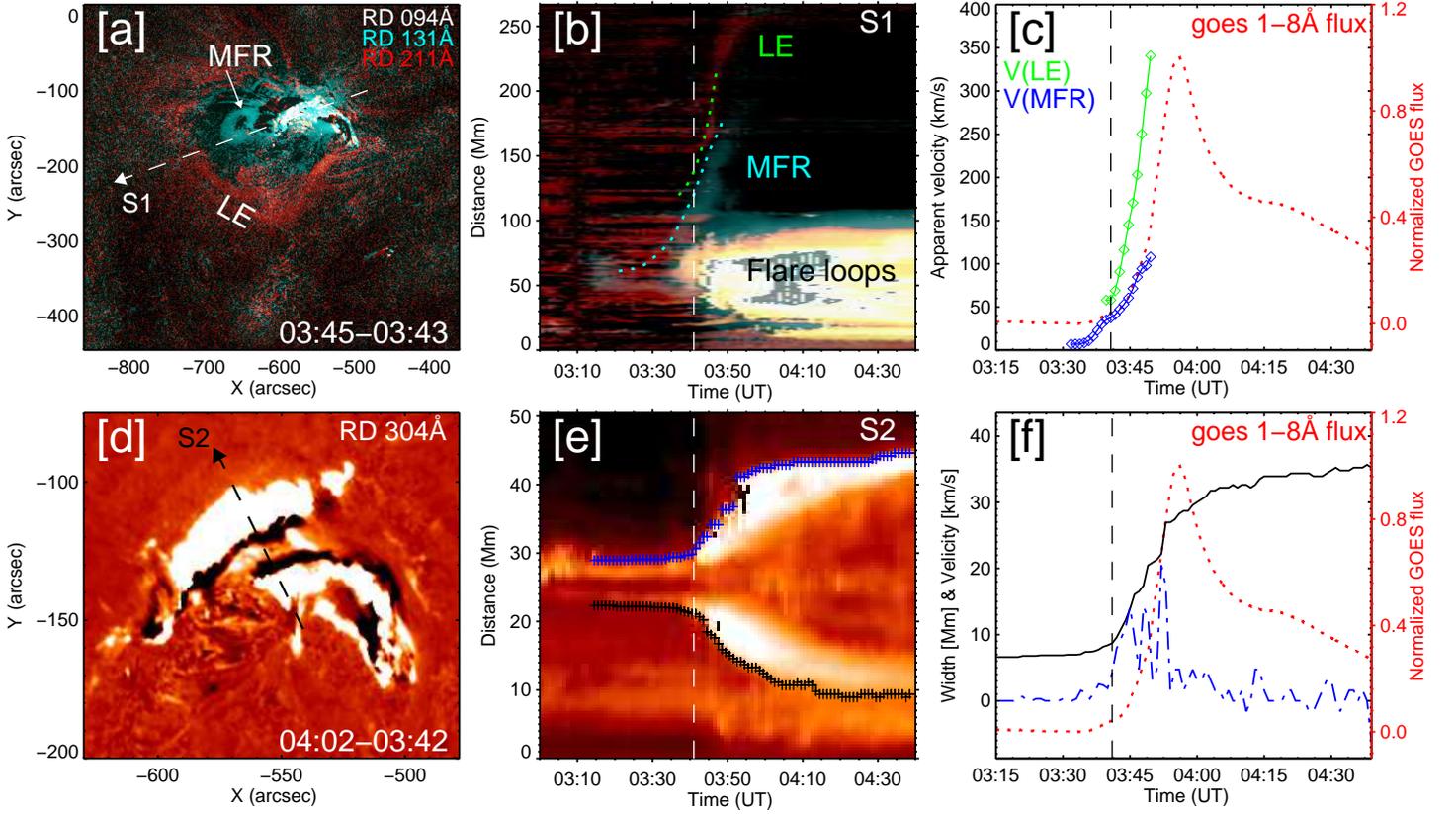}
              }
              \caption{(a): The multi-temperature structure of the nascent CME shown in a composite image for AIA 94, 131, and 211 \AA. (b): The height-time plot that made along the white dash line, $S1$, in panel (a), which shows the eruptive dynamic of the hot channel. (c): Temporal evolution of apparent velocity of leading front ($LF$) and the eruptive MFR, associated with $GOES$ flux. (d): The running-difference 304 \AA \ image shows the obvious FRs' lateral expansion after its formation. (e): The distance-time plot that made along black dash line, $S2$, displays the lateral expansion of FRs perpendicular to the PIL. Thereinto, each outer edges of two-ribbon flare are detected by edge detect technique. (f): Temporal evolution of two-ribbon flare's distance (plotted as black solid line) and its expansion velocity (plotted as blue dashed line), associated with $GOES$ flux. The vertical reference lines in panels (b-c) and (e-f) all denote the onset time of the hot channel eruption.}
   \label{fig4}
   \end{figure}
\begin{figure}[htb]
\centerline{
\includegraphics[width=.45\textwidth,clip=]{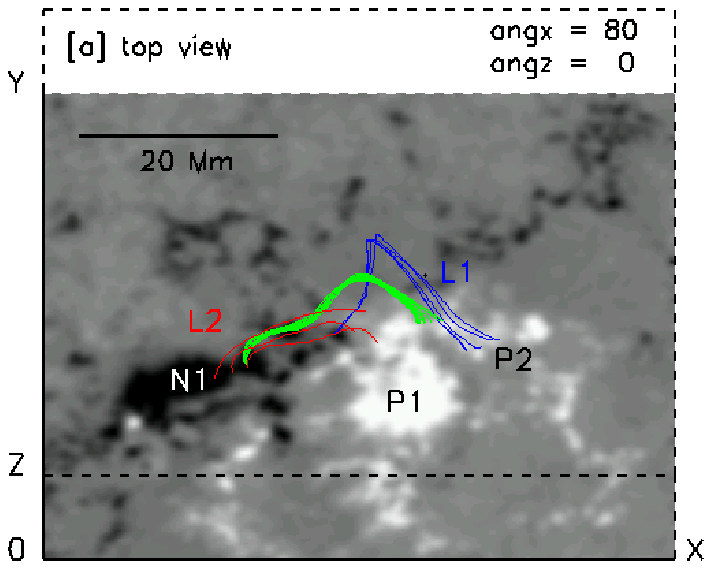}
\includegraphics[width=.45\textwidth,clip=]{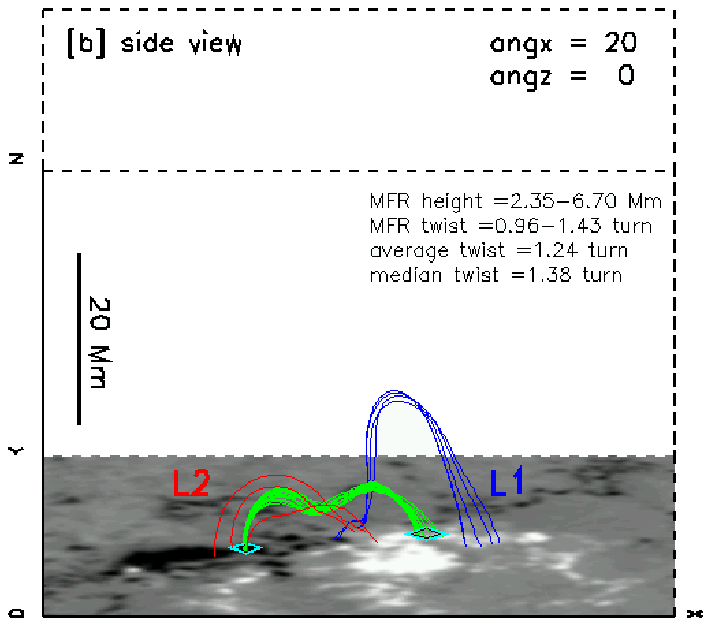}
          }
 \centerline{\includegraphics[width=1.1\textwidth,clip=]{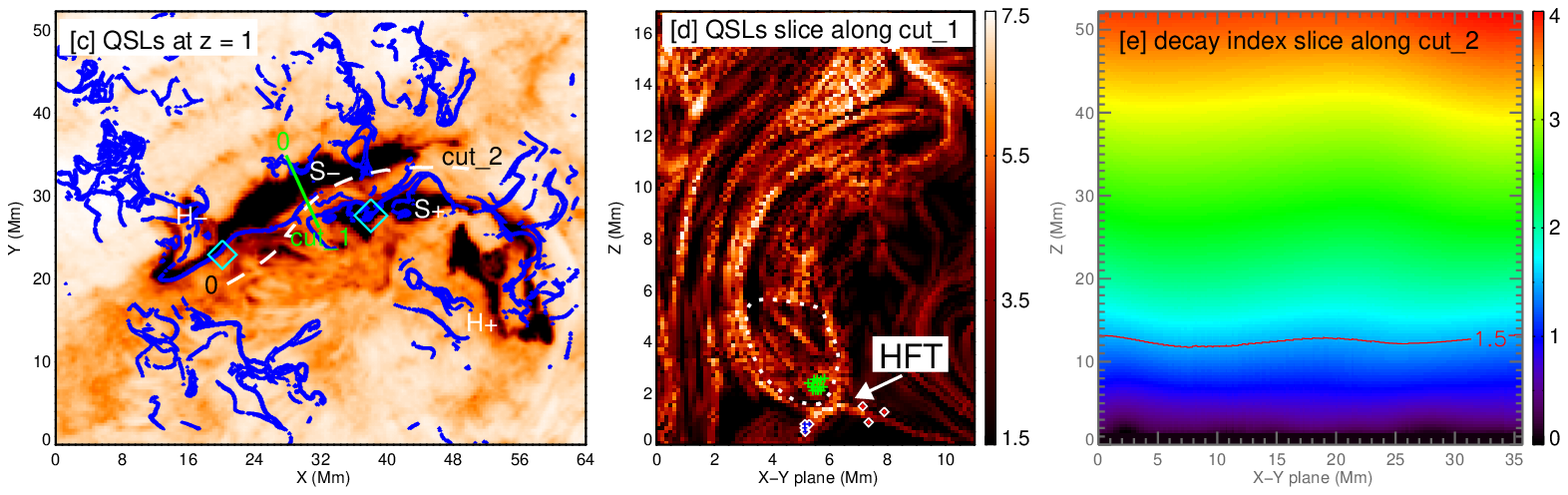}}
\caption{The pre-eruption magnetic configuration and its topology analysis at 03:12 UT. Panels (a) and (b) respectively show the pre-flare magnetic configuration seen from two different 3D perspective views, which performed within a box of $176\times144\times144$ uniformly
spaced grid points, corresponding to about $64\times52\times52$ Mm$^3$. Green lines trace the pre-existing seed MFR, while red/blue lines trace two sets of sheared arcades, $L1$ and $L2$, that hold the seed MFR. The basic information of the seed MFR is recorded in panel (b). Likewise, $P1$, $N1$, and $P2$ denote the photospheric flux as \nfig{fig8}(e). Panel (c): QSL map (with log(Q) $\geq $ 3.5) at Z = 1 are plotted as blue contours over the negative 304 \AA \ FR image. \textit{H+}, \textit{H-}, \textit{S+}, and \textit{S-} respectively mark the FR hooks or straight FR,according to their flux polarities. Cyan diamonds in panels (b) and (c) denote the extrapolated seed MFR's feet position. Panel (d): The vertical QSLs map cut (with log(Q) $\geq $ 1.5) along the green slice, \textit{cut1}, in panel (c), in which, the seed MFR, \textit{L1}, and \textit{L2} are located at green, blue, and red marks in cross section of Q-factor, respectively. Panel (e): The vertical decay index distribution along the white dash slice \textit{cut2} (also the east section of the PIL), in panel (c).}
\label{fig5}
\end{figure}
\begin{figure}[htb!]    
   \centerline{\includegraphics[width=0.6\textwidth,clip=]{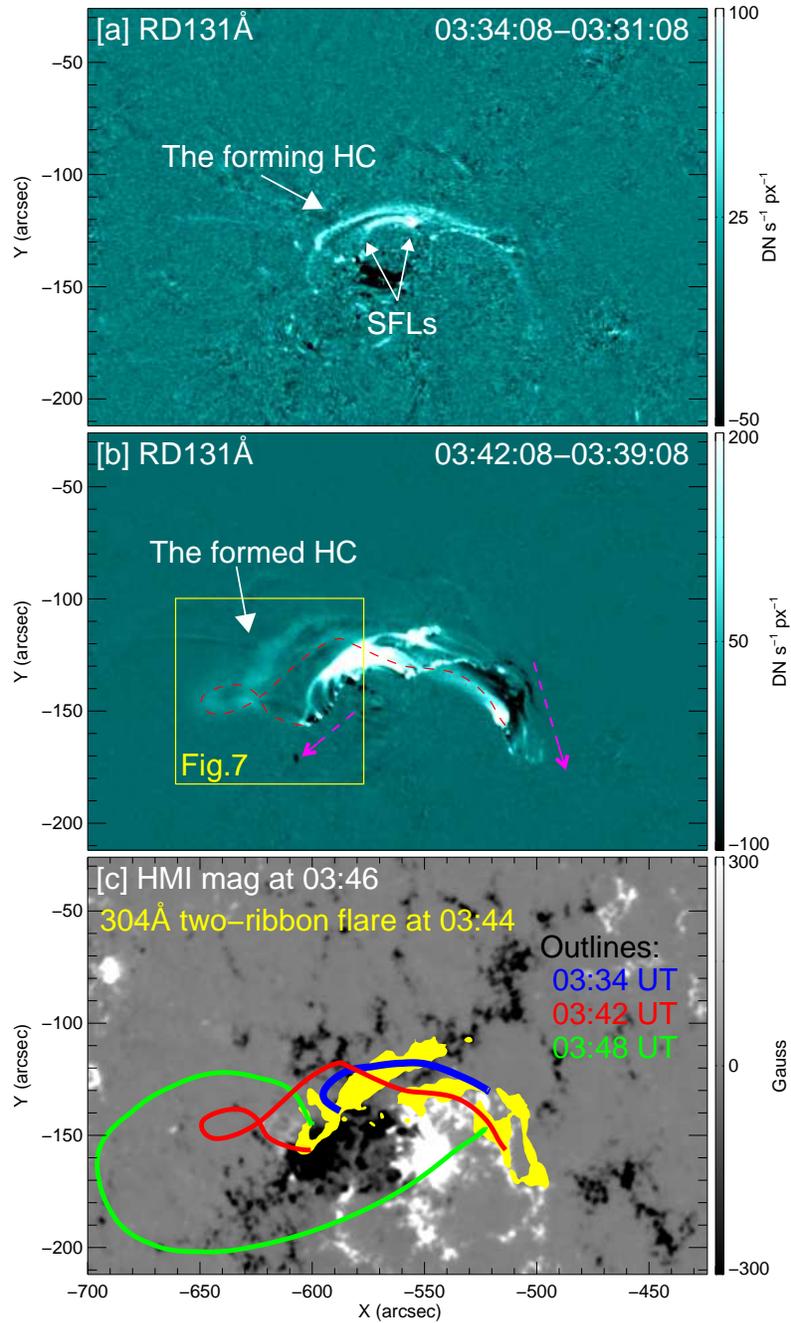}
              }
              \caption{The enlarged hot channel with drifting feet. Panels (a) and (b) respectively demonstrate the forming and formed hot channel at the running-difference 131 \AA \ images. In panel (b), red dash outlines the formed hot channel; purple dashed arrows indicate the apparent position drift of its feet; the yellow box denotes the FOV of \nfig{fig6}. Panel (c) presents the outlines of the hot channel at different times, which are related to the contour of two-ribbon flare and HMI magnetogram near 03:44 UT. }
   \label{fig6}
   \end{figure}
\begin{figure}[htb!]    
   \centerline{\includegraphics[width=1.0\textwidth,clip=]{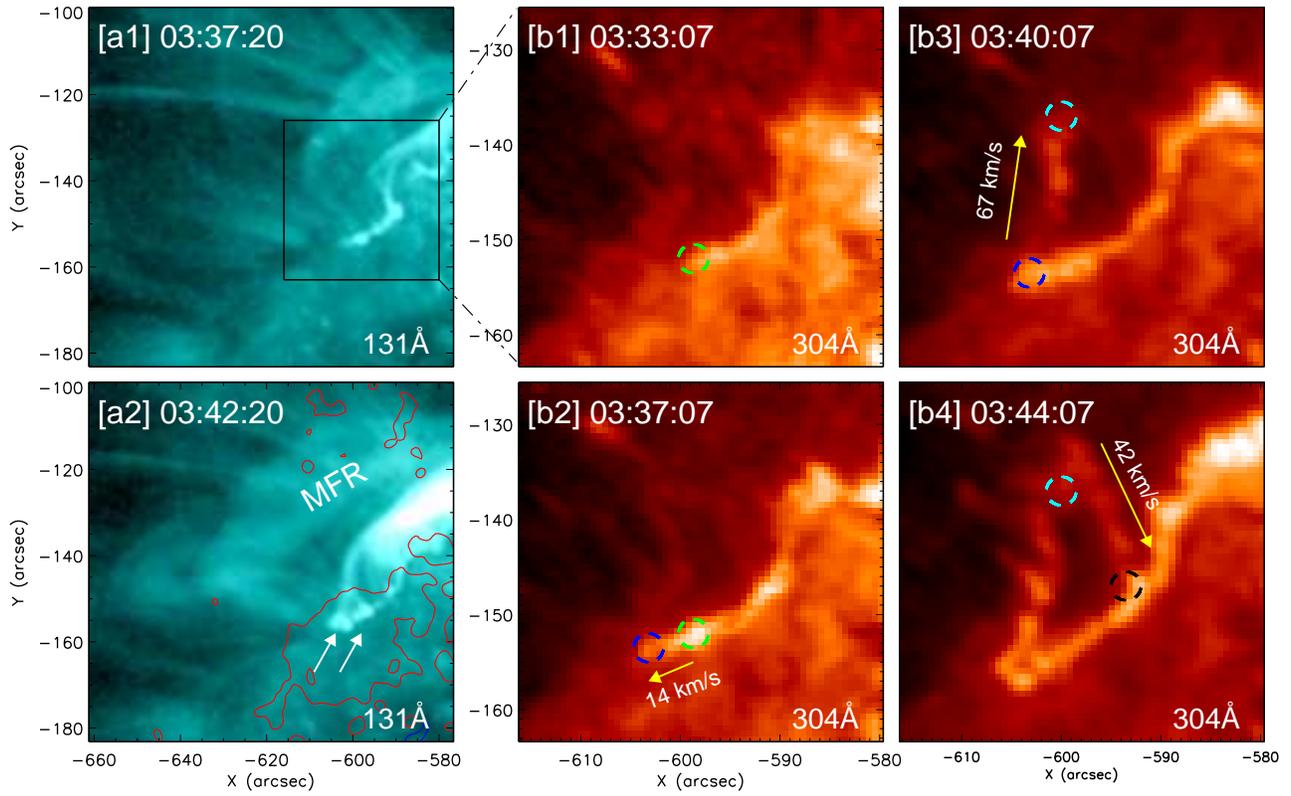}
              }
              \caption{Slipping reconnection signatures observed at the east feet of the forming hot channel. (a1-a2): Selected 131 \AA \ images illustrate the east feet of the hot channel, in which white arrows mark the flaring tiny knots. The red contour denotes the negative-polarity sunspot, which are saturated at $-$ 500 G. (b1-b4): Sequential 304 \AA \ images show the slipping flare kernels and its resulted closed FR hook. Thereinto, small circles with different color trace FR's slipping motion. }
   \label{fig7}
   \end{figure}
\begin{figure}[htb!]    
   \centerline{\includegraphics[width=0.6\textwidth,clip=]{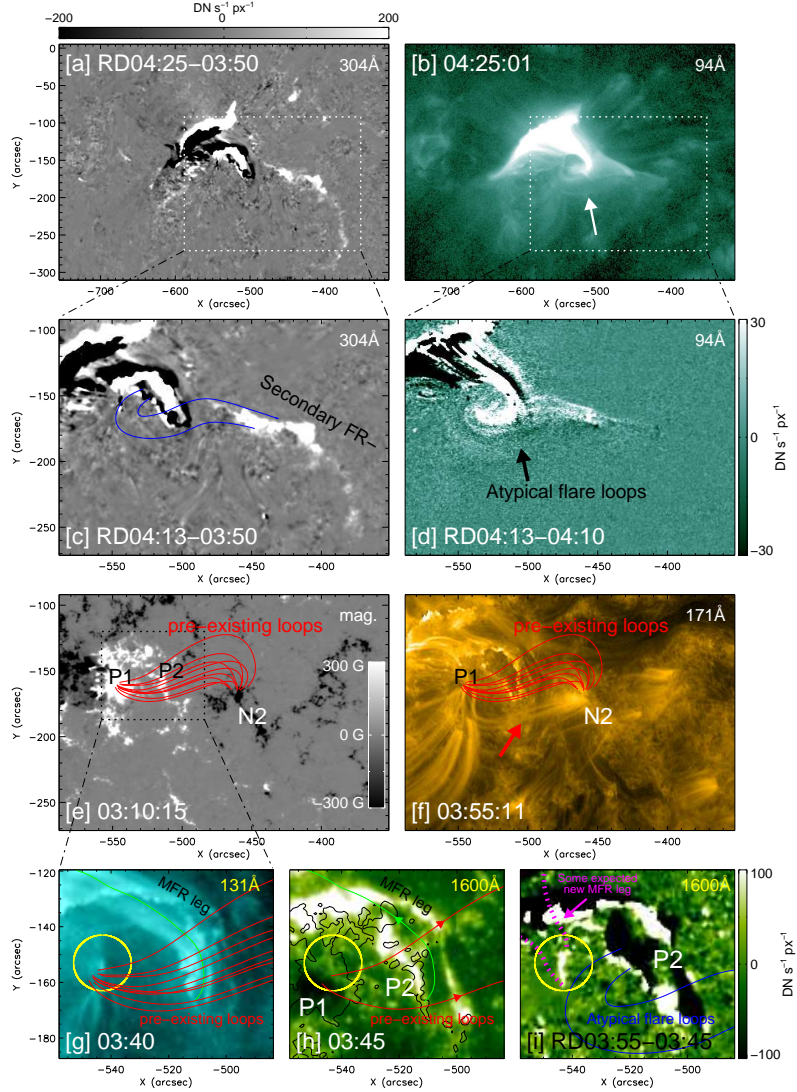}
              }
              \caption{Panels (a-d): the generation of the semi-circular secondary flare ribbon (FR) observed in running-difference 304 image and 94 image. Panels (e-f): HMI LOS magnetogram and 171 \AA \ image prior to the generation of secondary FR, in which the PFSS-extrapolated pre-existing loops are overlaid as red lines. Thereinto, $P2/N2$ denote the outer positive/negative fluxes; $P1$ denotes the positive flux of sunspot; Red arrow marks the pre-existing loops. Panels (g-i): The drifting signature of MFR's footpoints observed in 131 and running-difference 1600 \AA images. Thereinto, yellow circles enclose the end position of the pre-existing loops (red lines). Green and red lines in panels (g-h) represent the MFR's leg and pre-existing loops, while purple dash lines and blue solid lines in panel (i) outline the expected new MFR leg and the observed atypical loops. An animation of panels (a-d) is available. The animation covers 03:50 UT to 04:20 UT with 60 s cadence. The video duration is 3 s.}
   \label{fig8}
   \end{figure}
\begin{figure}[htb!]    
   \centerline{\includegraphics[width=0.7\textwidth,clip=]{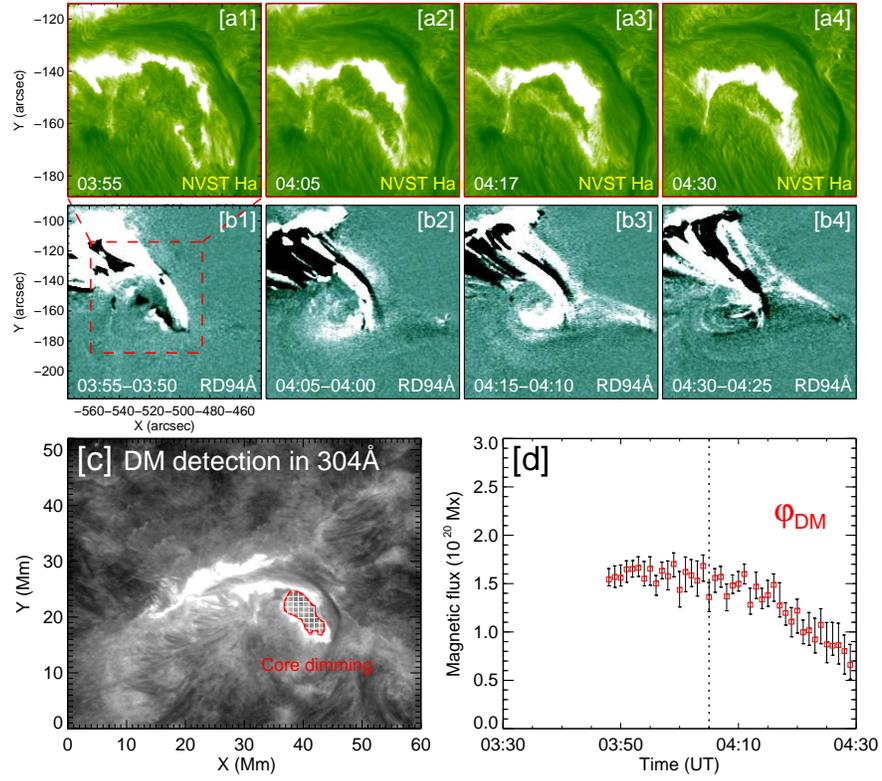}
              }
              \caption{Panels (a1-a4): Sequential NVST H$\alpha$ images show the dimming region's diminishment at the west feet of the erupting hot channel. Panels (b1-b4):  The appearance of atypical flare loops observed in running-difference 94 \AA \ images. Panels (c) and (d) illustrate the image detection of dimmed region in 304 \AA \ image, and the temporal variation of the magnetic flux ($\Phi_{DM}$) that computed over the west core dimming region. The vertical dash line indicates the moment (04:05 UT) that the atypical flare loops obviously formed.}
   \label{fig9}
   \end{figure}

One striking feature in the present study  is that the two-ribbon flare underwent a clear elongation-to-expansion dynamics evolution. To our knowledge, similar two-phase FR dynamics have been documented by several previous literatures \citep{1990SoPh..125..321K,2009ApJ...692.1110Q}, which has been suggested as an indicator of a macroscopically two-phase reconnection \citep{2009ApJ...692.1110Q}. Compared with these previous works, in our study, direct physical links among the elongation-to-expansion FR dynamics and its related eruptive coronal MFR are unambiguously captured. Our analysis revealed that the hot-channel-like MFR was rapidly built up via tether-cutting type 3D pre-eruption reconnection in the formation course of the two-ribbon flare, during which sporadic flare kernels appeared and elongated along two sides of the PIL where QSLs distributed; while the perpendicular expansion of the two-ribbon flare was subsequently caused by a quasi-2D ``arcade reconnection''. Note that this 3D pre-flare reconnection was triggered near the HFT, thus it often be treated as a precursor slipping reconnection by some researchers \citep{2016ApJ...830..152L}.
Moreover, as shown in \nfig{fig4}, the elongation and expansion motion of two-ribbon flare are well divided by the exponential acceleration of the rising hot channel. In accord with the extended 3D standard eruption model, these features imply that a possible involvement of ideal MHD instability in the solar eruption may lead the preceding 3D reconnection near the HFT to a quasi-2D arcades reconnection with increasing height \citep{2010ApJ...708..314A}, naturally causing such an elongation-to-expansion dynamics transition in its two-ribbon flare.
\par
Identifying the footpoints of erupting CME flux-ropes is a crucial step to bridge interplanetary CME flux-ropes down to its solar progenitor. Although identifying the footpoints of CME flux-rope during the instantaneous eruptive process is quite difficult, core dimmings are thought to map the feet of the erupting CME flux-rope \citep{2018ApJ...857...62V,2019ApJ...879...85V}, since the magnetic field there can open to interplanetary space allowing plasma to escape outwards \citep{1997ApJ...491L..55S,2000JGR...10527251W,2007ApJ...662L.131J}. This conjecture has also been indirectly supported by some 3D flux-rope models, in which footpoints of CME flux-ropes are predicted to be enclosed by J-shaped or closed flare hooks \citep{1996JGR...101.7631D,2012A&A...541A..78P,2013A&A...555A..77J}. Accordingly, applying this conjecture, several observational works further proposed that the amount of CME mass that supplied over coroanl dimmings  \citep[e.g.,][]{2019ApJ...879...85V} and the poloidal flux of CME flux-ropes \citep[e.g.,][]{2014ApJ...793...53H} may be estimated through identifying core dimmings with time.
However, under the framework of 3D extensions to the standard eruption model, \citet{2012A&A...543A.110A} finds that the line-tied footpoints of the CME flux-rope can dynamically drift to new position due to a new reconnection geometry among the erupting flux-rope and thereby inclined arcades, namely ``ar-fr" reconnection. More recently, the predicted footpoint drift of erupting CME flux-ropes is now being confirmed by new observations \citep{2019arXiv190802082Z,2019arXiv190903825L}.
\par
In the present study, obvious MFR¡¯s footpoint drift signatures are also detected near two conjugate core dimmings. Thereinto, the drift of its east footpoints manifested as an apparent slipping motion of flare
kernels, resulting in the formation of a closed FR hook.
In particular, an extra 3D reconnection process is also been observed near the west core dimming, which induced the generation of a secondary flare ribbon and a rapid diminishment of the west core dimming at the same time. It is worthy to note that this reconnection geometry took place among the west leg of the erupting MFR and thereby pre-existing inclined arcades, which is analogous with the configuration that required in the so-called ``ar-fr" reconnection. As a result of this reconnection geometry, it is likely that the erupting MFR partly drifted its west leg to a new position with a concurrent footpoint brightening signature in 1600 \AA \ images, while the pre-existing inclined arcades finally turned into a set of atypical flare loops connecting into the west core dimming. Consequently, due to the dynamic footpoint drift of the erupting MFR, the integral magnetic flux over the west core dimming soon decreased from $1.70\times 10^{20}$ to $0.65\times 10^{20}$ Mx within 40 minutes. This result evidences that the eruptive MFR's footpoint drift can cause a rapid and pronounced diminishment of core dimmings in some events. In that case, the poloidal flux and plasma supply of the erupting CME flux-rope that computed through identifying such decreasing core dimmings during eruptions may suffer a nonnegligible underestimation. These results strongly reveal that, once the erupting MFR reconnects with thereby magnetic field, mapping its real
footpoints down to the Sun's surface would become more difficult than previously thought.
\par
As the last remark, our observation also shed a new light on the generation of the so-called secondary flare ribbon.
Previous observations generally contribute their generations to the existence of a complex flux distribution \citep{2016ApJ...823..136Z} or a null-point configuration \citep{2014ApJ...781L..23W,2015ApJ...812...50J,2017ApJ...845...26J} above the main eruption. In our case, a semi-circular secondary FR was also found to generate not far from the eruption source region after the flare peak time. However, our analysis reveal that the generation of this newborn secondary FR was directly caused by an new 3D reconnection among the inflating west feet of the eruptive MFR and a set of nearby inclined arcades, instead of the pre-existence of a large-scale fan-spine configuration. This implies that at least in some cases, the incidental 3D reconnection among the erupting MFR and thereby arcades can also be regarded as an alternative interpretation for the generation of secondary FR.
\par

\acknowledgments
We sincerely thank the anonymous referee for valuable comments and suggestions that improve our manuscript. We also thank Yi Bi and Haidong Li for helpful discussions, as well as Xiaoli Yan for useful suggestions. The data used here are courtesy of the SDO, NVST, SOHO, and GOES science teams. As one of the primary observing facilities of the \textit{Fuxian Solar Observatory} \footnote{http://fso.ynao.ac.cn/index.aspx}, NVST is jointly operated and administrated by Yunnan
Observatories and Center for Astronomical Mega-Science,
Chinese Academy of Science. This work is supported by the Natural Science Foundation of China (NSFC) under grants 11633008, 11703084, 11873088, and 11973085. K. F. J is supported by NSFC under grants 11773072 and 11873027. J. Y. Y is also supported by NSFC under grant 11933009 and by Key Programs of the Chinese Academy of Sciences (QYZDJ-SSW-SLH012).

\end{document}